\documentclass[12pt]{article}
\usepackage{amsmath,amssymb,amsfonts}
\usepackage[paper=letterpaper,margin=1.0in]{geometry}
\usepackage{graphicx}
\usepackage{bm}
\usepackage{color}

\newcommand{\red}[1]{{\color{red} #1}}
\newcommand{\blue}[1]{{\color{blue} #1}}
\newcommand{\cyan}[1]{{\color{cyan} #1}}

\begin{document}

%\rightline{\footnotesize Preprint Number}

\begin{center}

\vskip 1.0cm
\centerline{\Large\bf 
Universal Acceleration and Fuzzy Dark Matter
}
\vskip 1.0cm

\renewcommand{\thefootnote}{\fnsymbol{footnote}}
\centerline{\bf
Douglas Edmonds${}^{1}$\footnote{edmonds@psu.edu (Corresponding Author)},
Joshua Erlich${}^{2}$\footnote{jxerli@wm.edu},
Djordje Minic${}^{3}$\footnote{dminic@vt.edu} and
Tatsu Takeuchi${}^{3}$\footnote{takeuchi@vt.edu}
}
\renewcommand{\thefootnote}{\arabic{footnote}}

%\vskip 0.5cm

{\it
${}^1$Department of Physics, Penn State Hazleton, Hazleton, PA 18202 USA\\
%\vskip 0.2cm
${}^2$Department of Physics, William and Mary, Williamsburg, VA 23185 USA\\
%\vskip 0.2cm
${}^3$Department of Physics, Virginia Tech, Blacksburg, VA 24061 USA
}

\vskip 0.5cm
\centerline{Submission date: March 30, 2024}
\vskip 0.5cm

\begin{abstract}
Observations of velocity dispersions of galactic structures over a wide range of scales point to the existence of a universal acceleration scale $a_0\sim 10^{-10}$ m/s$^2$. Focusing on the fuzzy dark matter paradigm, which proposes ultralight dark matter with mass around $10^{-22}$ eV and de Broglie wavelength $\lambda\sim {\rm few}\times10^{2}$ parsecs, we highlight the emergence of the observed acceleration scale from quantum effects in a fluid-like description of the dark matter dynamics. We then suggest
the possibility of a natural connection between the acceleration scale and dark energy within the same paradigm.

\end{abstract}

\vspace{0.5cm}

\begin{center}
This paper has been awarded 3rd Prize in the\\Gravity Research Foundation 2024 Awards for Essays on Gravitation.
\end{center}

\end{center}

%%%%%%%%%%%%%%%%%%%%%%%%%%%%%%%%%%%%%%%%%%%%%%%%%%%%%%%%%%%%%%%%%%%%%
\newpage
%\section{Introduction}

The question of whether the observed universe requires
non-baryonic matter beyond the Standard Model, or whether the fundamental description
of gravity should be modified beyond the realm of general relativity is one of the outstanding questions in contemporary physics.

In this essay, we highlight the evidence for a universal acceleration scale, $a_0\sim 10^{-10}\,\mathrm{m/s^2}$, in the observed
structures in the universe, and examine whether the existence of such a scale
is compatible with the fuzzy dark matter paradigm \cite{Hu:2000ke,Hui:2016ltb}.
%The answer offered in the following is that indeed a fundamental acceleration set by the Hubble scale \cite{Milgrom:1998sy} %(which suggests a relation to another fundamental puzzle of physics, that of dark energy) 
%is found at all observed scales of structures in the universe and that this fundamental scale is consistent with the fuzzy dark matter paradigm responsible for the emergence of the observed structures. %(galaxies, clusters, etc). 
In particular, we discuss a possible relationship between the acceleration scale and the
quantum pressure of fuzzy dark matter models \cite{Lee:2019ums}, which also suggests that the acceleration scale may change with redshift.
We further conjecture a relationship between the quantum pressure and dark energy, which could explain why $a_0\sim cH_0/(2 \pi)$ \cite{Milgrom:1998sy}.

The empirically established Tully-Fisher (TFR) \cite{Tully:1977fu} and Faber-Jackson (FJR) \cite{Faber:1976sn} relations %(discussed below) 
indicate that the observed velocity dispersions of galaxies and clusters of galaxies can be related to the baryonic content of those structures, independently of the dark matter content.
The TFR and FJR in turn suggest a fundamental
acceleration scale $a_0$, and given the relevance of baryonic matter
in these scaling relations, provide motivation for 
a modification of gravity
and/or modification of inertia \cite{Milgrom:1983ca,Milgrom:1983pn,Milgrom:1983zz}.
However, it is difficult to reconcile such attempts %at modification of gravity and/or modifications of inertia 
with the observed data at all scales, including the precision data from the early universe provided by the cosmic microwave background (CMB).
On the other hand, in traditional cold dark matter (CDM) models the emergence of an acceleration scale in galactic structures is not currently understood, although simulations of CDM models hint at such behavior \cite{1998MNRAS.295..319M,10.1093/mnras/stw2461,10.1093/mnras/stw2691}.

The TFR is an empirical correlation between the absolute luminosity $L$ of a spiral galaxy and the velocity $v_\infty$ of stars and gas in the galaxy's outskirts as determined from the width of the 21 cm HI line,
\begin{equation}
L \;\propto\; v_\infty^n\;,
\end{equation}
where $n \approx 4$. The TFR is often used as a distance indicator: the absolute luminosity can be determined from the observed velocity, and then compared to the apparent luminosity. Comparison of the absolute and apparent luminosities then yields an estimate of the distance to the galaxy.

An alternative correlation is the baryonic Tully-Fisher relation (BTFR). In the BTFR, baryonic mass is used in place of absolute luminosity:
\begin{equation}
M_\mathrm{bar}\;\propto\; v_\infty^n\;,
\label{BTFR}
\end{equation}
where, again, $n \approx 4$. The BTFR implies a universal acceleration scale for rotationally supported systems (high surface brightness (HSB), low surface brightness (LSB), irregular, and dwarf irregular galaxies):
\begin{equation}
a_0 \;=\; \dfrac{v_\infty^4}{GM_\textrm{bar}}\;,
\end{equation}
where, $G$ is Newton's gravitational constant. Analysis of galactic rotation curve data in Ref.~\cite{McGaugh:2011ac} yields $a_0=(1.3\pm 0.3)\,\times 10^{-10} \mathrm{m/s^2}$.

There is an analogous correlation observed for pressure supported systems (dwarf spheroidals, elliptical galaxies, galaxy clusters). The FJR is an empirical correlation between the absolute luminosity and the line-of-sight velocity dispersion, $\sigma$:
\begin{equation}
L \;\propto\; \sigma^n\;.
\end{equation}
Original observations \cite{Faber:1976sn} indicated $n \approx 4$. However, subsequent observations found a spread in values of $n$. We typically find values in the range $n = 3 \sim 5$. The measured value depends on the dataset and the methods used, and $n$ can be as low as 2 for dwarf galaxies. Interestingly, globular clusters, which are thought to have formed without dark matter halos (though this is an open question) follow the FJR, though there is considerable scatter.

Just as with the TFR, an alternative to the FJR is the baryonic Faber-Jackson relation (BFJR), where the baryonic mass replaces the absolute luminosity:
\begin{equation}
M_\textrm{bar} \;\propto\; \sigma^n\;.
\end{equation}
For $n=4$, this implies a universal acceleration scale for pressure supported systems:
\begin{equation}
a_0 \;=\; \dfrac{\sigma^4}{G M_{\textrm{bar}}}\;.
\label{eqn:a_elliptical}
\end{equation}
Interestingly, the acceleration scale measured for pressure supported systems is very similar to the one found for rotationally supported systems. For example, Ref.~\cite{Chae_2020} finds a value of $a_0 = (1.5_{-0.6}^{+0.9})\,\times 10^{-10} \mathrm{m/s^2} $ for slow-rotator E0 galaxies. 
 
The data presented in Fig. 1 for various structures over a wide range of scales provide compelling evidence for such an acceleration. 
We define the ``virial acceleration" to be $\sigma^4/GM_\text{bar}$, where $\sigma$ is the velocity dispersion, $G$ is Newton's gravitational constant, and $M_\text{bar}$ is the total baryonic mass of the structure. In Fig. 1, we plot the virial acceleration for structures across observed scales, from globular clusters to galaxy clusters. This is a mass range of $\sim 10^{4-14} M_\odot.$ The central values for various observed structures are extraordinarily similar, though the spread in values differ for different structures.

\begin{figure}[t]
\begin{center}
\includegraphics[width = 0.8\textwidth]{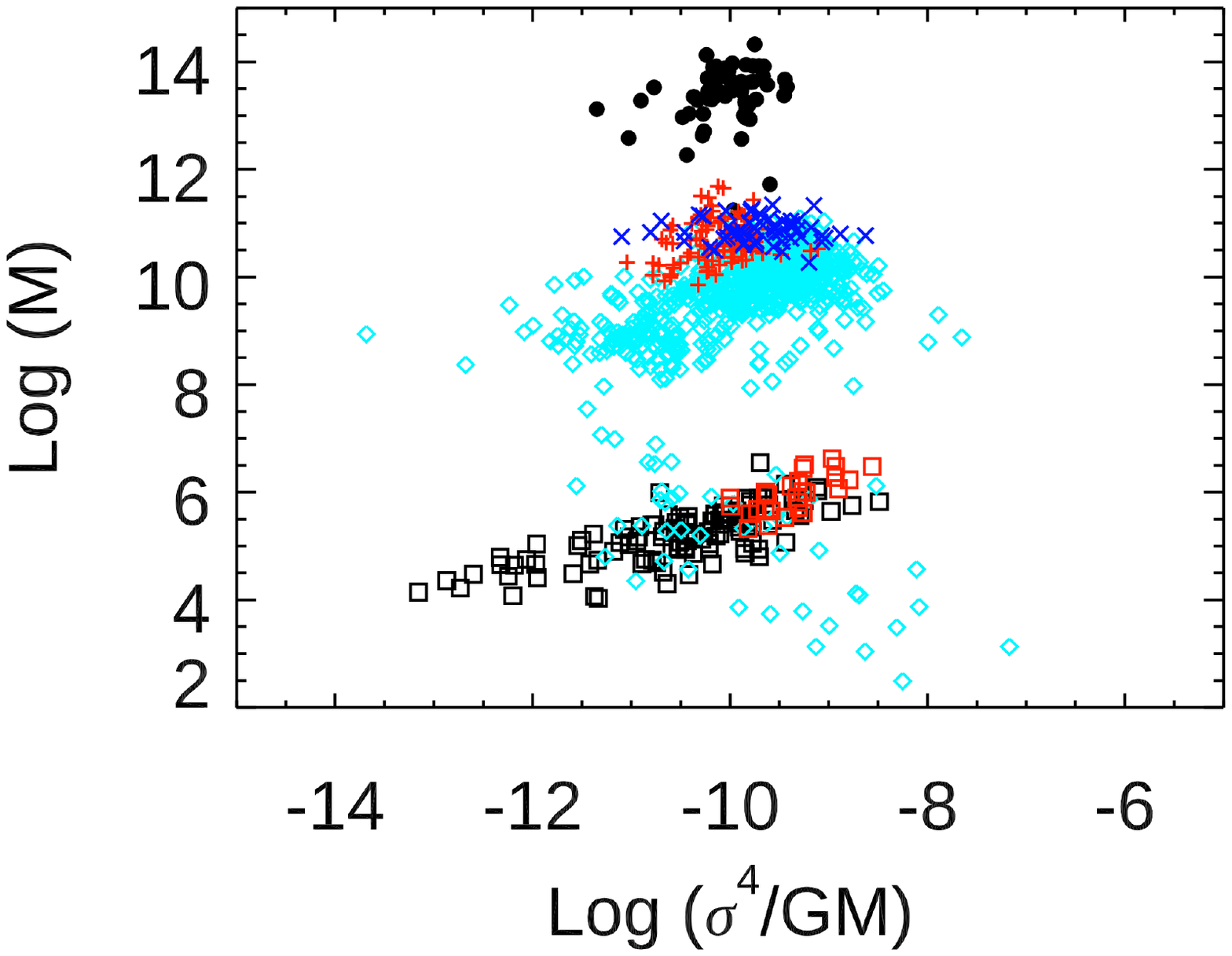}
\label{snowman}
\caption{The virial acceleration for structures spanning about 10 decades of scale $M/M_\odot = 10^{4\sim 14}$. 
The data points are: $\bullet$ galaxy clusters \cite{Zhang:2010qk},
{\scalebox{0.8}{$\red{\bm{+}}$}} elliptical galaxies \cite{Cappellari:2012ad,Cappellari:2012ae},
{\scalebox{0.8}{$\blue{\bm{\times}}$}} elliptical galaxies \cite{Belli:2013oia,Belli:2014yfa},
$\cyan{\bm{\diamond}}$ elliptical, dwarf elliptical, and dwarf spheroidal galaxies \cite{stw1248},
{\scalebox{0.7}{$\bm{\square}$}} Milky Way globular clusters \cite{Baumgardt:2018},
{\scalebox{0.7}{$\red{\bm{\square}}$}} M31 (Andromeda) globular clusters \cite{Strader:2009hg}.
}
\end{center}
\end{figure}
 
Observations of high-redshift galaxies may offer additional insights. Currently, different models predict different redshift dependence for the observed acceleration scale. For example, CDM simulations indicate that the acceleration scale increases with redshift, though the amount of change depends on details of the feedback implementation. Current observations do not include high enough redshift to address this issue definitively, but going out to $z \sim 2.5$, we see no evidence for a redshift dependence. In Fig. 2, we plot the Radial Acceleration Relation \cite{McGaugh_2016} (RAR) for a sample of 100 high-$z$ galaxies, with redshifts in the range $z = 0.5\sim 2.5$. As we see in the plot, the current data are consistent with no change in acceleration scale with redshift. However, we note that future observations are needed at much higher redshifts to see if this continues to hold.

\begin{figure}[t]
\centering
\includegraphics[width = 0.8\textwidth]{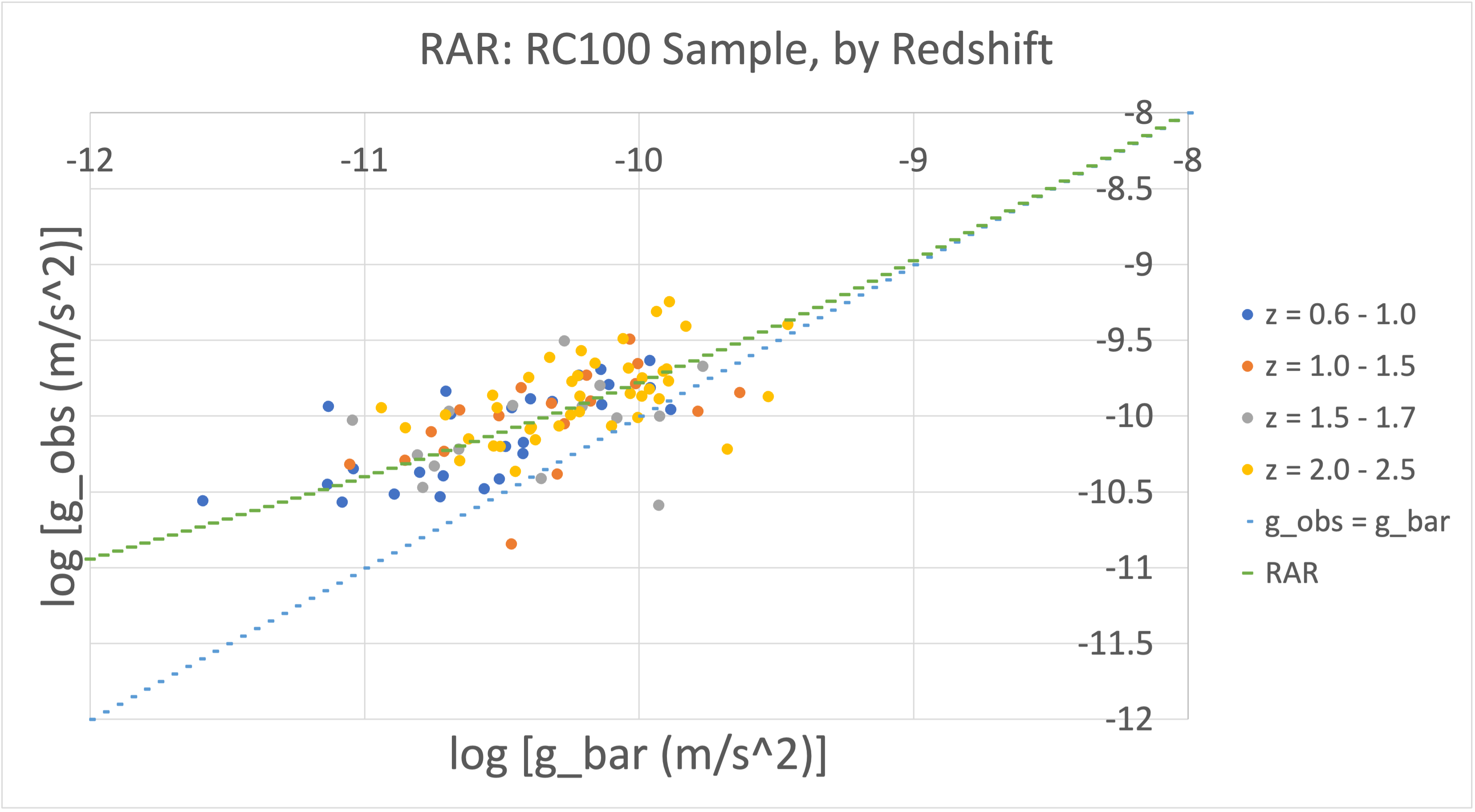}
\label{highz}
\caption{The Radial Acceleration Relation (RAR) for a sample of 100 high-$z$ galaxies in the RC100 sample. Here, $g_{obs}$ and $g_{bar}$ are the observed acceleration and the acceleration due to baryons only, respectively. Galaxies plotted are in the redshift range $z \approx 0.5-2.5$.} 
\end{figure}

Note the following numerological connection between $a_0$ and the
Hubble constant $H_0$, to wit, $a_0 \sim c H_0/(2 \pi)$, where $c$ is
the speed of light.
Why would the Hubble scale have anything to do with structure formation?
Fuzzy dark matter models suggest an origin for $a_0$ and possible explanations for these related features.

The fuzzy dark matter (FDM) paradigm \cite{Hu:2000ke,Hui:2016ltb} is interesting as it presents solutions to problems encountered in canonical dark matter models regarding structure formation. Fuzzy dark matter models contain ultralight $(10^{-22}\ {\rm eV})$ particles, perhaps axion-like, with de Broglie wavelength around $(10^2-10^3)$ parsecs, long enough to be relevant for structures at galactic core scales. In this paradigm, the dark matter may be treated as a nonrelativistic quantum fluid. For this purpose, the Schr\"odinger equation is written in its Madelung form \cite{Madelung:1927ksh} in terms of the magnitude and phase of the wavefunction, $\psi=\sqrt{\rho}\,e^{iS}$, adapted to the background of an expanding universe \cite{Hui:2016ltb}:
\begin{eqnarray}
\frac{\partial\rho}{\partial t}+3H\rho+\frac{1}{a}\,\nabla\cdot(\rho\mathbf{v}) & = & 0\;, \label{eq:Madelung1}\\
\frac{\partial\mathbf{v}}{\partial t}+H\mathbf{v}+\frac{1}{a}\,(\mathbf{v}\cdot\nabla)\mathbf{v} & = & -\frac{1}{a}\,\nabla\Phi
+\underbrace{\frac{\hbar^2}{2a^3m^2}\nabla\left(\frac{\nabla^2\sqrt{\rho}}{\sqrt{\rho}}\right)}_{\displaystyle V_Q}\;,
\label{eq:Madelung}
\end{eqnarray}
where $a(t)$ is the scale factor, $H=\dot{a}/a$, $\Phi$ is the gravitational potential, and $V_Q$ is the quantum potential term.
The fluid velocity is given by 
\begin{equation}
    \mathbf{v}\;=\;\frac{\hbar}{ma}\,\nabla S\;.
\end{equation}
The probability density function $\rho$ is identified with the physical dark-matter density, suitably normalized.

The quantum potential $V_Q$ has dimensions of acceleration. This term is responsible for suppressing the formation of cuspy structures in the galactic core and satellites in the galactic halo \cite{Hui:2016ltb}. Replacing the spatial derivatives by the inverse of the de Broglie wavelength $\lambda_{{\rm dB}}\sim 300$ pc, and the mass with $m\sim 10^{-22}$ eV, the typical acceleration associated with the quantum potential is 
$a_\text{FDM}\sim 2\times 10^{-10}$ m/s$^2$. This is an interesting acceleration scale, of the same order as the acceleration scale observed in the Tully-Fisher and Faber-Jackson relations. It is then tempting to attempt to explain the virial acceleration of galactic structures as due to the quantum pressure in fuzzy dark matter, and indeed this connection was explored in Ref.~\cite{Lee:2019ums}. There it was found that demanding an equilibrium between the gravitational potential and the quantum potential for the fuzzy dark matter leads to the observed acceleration scale, with the gravitational potential providing a link between the baryonic and dark matter, which ultimately leads to the Tully-Fisher relation. We note that, in the theory, as long as the density of
baryonic matter falls off more rapidly than the dark matter density,
dark matter will start to dominate at the acceleration scale set by the FDM quantum potential. This agrees with observations which indicate that $a_0$ is the acceleration at the radius where dark matter starts to dominate.

Now we make a suggestion regarding the origin of the fundamental
acceleration $a_0$ in the context of FDM models:
In a field-theoretic context the quantum potential in the generalization of Eqs.~(\ref{eq:Madelung1}) and (\ref{eq:Madelung}) naturally generates the 
vacuum energy, once one uses the vacuum wave functional in its expression. For example, for a free and massless bosonic field 
(Ref.~\cite{Undivided_Universe}, page 244-245), the vacuum 
wave functional gives the leading term in the 
quantum potential $(V_Q)_0 = \sum_{\mathbf{k}} \frac{1}{2} k$, which is the standard expression
for the vacuum energy that is quartically divergent in $3+1$ dimensional spacetime.
This expression can be related to the observed value for the cosmological constant by a careful computation that uses the 
relation between the vacuum energy and the volume of phase space,
together with the Bekenstein bound on the gravitational entropy \cite{Freidel:2022ryr,Berglund:2022qsb}. This in turn gives the relation between the vacuum energy scale ($10^{-3}$ eV) and 
the geometric mean of the Hubble scale ($10^{-34}$ eV) and the Planck scale ($10^{28}$ eV, set by Newton's gravitational constant).
Then the above leading vacuum contribution to the quantum potential gives an analogous contribution to the gravitational potential (and thus acceleration) for structures defined by the balancing of the quantum and gravitational potentials.

Thus a universal acceleration stemming from the leading vacuum term in the quantum potential of FDM may be related to the Hubble scale (as required by the empirical relation $a_0 \sim c H_0/(2 \pi)$), which would make it appear to be fundamental. Note that applying the same reasoning
(see for example, \cite{Berglund:2023gur}) that gives the scale of vacuum energy as a geometric mean of an infrared (IR) and an ultraviolet (UV) scale, to the Hubble scale and the vacuum energy scale itself, one can obtain the scale of $10^{-19}$ eV, which is not that far from the energy scale of FDM quanta ($10^{-22}$ eV) needed to reproduce the fundamental acceleration directly from the quantum potential. The scale of $10^{-19}$ eV  
could be associated with the heavier soliton degrees of freedom of FDM that are responsible for the structure of FDM halos, and thus
it is relevant for structure formulation. Hence, the vacuum contribution and the leading FDM contribution could be related as they stem from the same vacuum wave functional.  Similarly, by self-consistency of the above equations of motion, baryons share the same universal acceleration, because of the fact that they move in a self-consistent gravitational potential that is balanced by the quantum potential.
Also, given the role of the quantum potential in the FDM structure formation %(solitons, granular structure, quantum interference) 
it may be possible to relate structure formation to the Hubble scale on all scales and all structures, as indicated by the observation summarized in Fig. 1.

It is worth noting that both the Hubble scale and the quantum potential term in Eq.~(\ref{eq:Madelung}) depend on the
redshift $z$, and thus the fundamental acceleration can be expected to be $z$ dependent. However, as Fig. 2 indicates, the observed data, available only for redshifts up to $z\sim 2.5$, do not suggest
any $z$ dependence for $a_0$. Therefore, studies of the redshift dependence of $a_0$ at higher $z$ could provide a test of these ideas.

\bigskip
\noindent
\textbf{Acknowledgments:}
DM and TT are supported in part by the US Department of Energy (DE-SC0020262).
DE, DM, and TT are also supported by the Julian Schwinger Foundation.
JE is supported in part by a Plumeri Award for Faculty Excellence at William \& Mary.

%%%%%%%%%%%%%%%%%%%%%%%%%%%%%%%%%%%%%%%%%%%%%%%%%%%%%%%%%%%%%%%%%%%%%

\bibliographystyle{unsrt}
\bibliography{Fuzzy}

\end{document}